\documentclass[12pt]{iopart}

\usepackage{iopams}
\usepackage{braket}
\usepackage{amsfonts}
\usepackage{color}
\usepackage[dvipsnames]{xcolor}
\usepackage{amssymb}
\usepackage{mathrsfs}
\usepackage{graphicx}
\usepackage{lmodern}
\usepackage{lineno}
\usepackage{hyperref}
\usepackage[normalem]{ulem}

\newcommand{\beq}{\begin{eqnarray}}
\newcommand{\eeq}{\end{eqnarray}}

\def\keywords#1{\vspace{10pt}
     \begin{indented}
     \item[]\rm Keywords: #1\par
     \end{indented}}

\begin{document}



\title{The Polymer representation for the scalar field: A Wigner functional approach}
\author{Jasel Berra--Montiel$^{1,2}$}

\address{$^{1}$ Facultad de Ciencias, Universidad Aut\'onoma de San Luis 
Potos\'{\i} \\
Campus Pedregal, Av. Parque Chapultepec 1610, Col. Privadas del Pedregal, San Luis Potos\'i, SLP, 78217, Mexico}
\address{$^2$ Dual CP Institute of High Energy Physics, Mexico}

\eads{\mailto{\textcolor{blue}{jasel.berra@uaslp.mx}}}


\begin{abstract}

In this paper, we analyze the polymer representation of the real--valued scalar field theory within the deformation quantization formalism. Specifically, we obtain the polymer Wigner functional by taking the limit of Gaussian measures in the Schr\"odinger representation. The limiting functional corresponds to the polymer representation derived by using algebraic methods such as the GNS construction, and the Fock quantization procedure.
\end{abstract}


\section{Introduction}
One of the main open problems in fundamental physics is related to study the behaviour of quantum fields in a completely background--independet manner. The term backgroun-independent,  broadly means that a theory requires to be only defined on a manifold endowed with no geometrical and metric structure. A notable examples of this situation stand out the early stages of the universe, the structure of the space--time at short distances and black holes evaporation. According to Loop Quantum Gravity (LQG) \cite{RovelliLQG},\cite{Thiemann},\cite{Lewandowski}, a background--independent and non--perturbative approach based on quantum geometry, in order to obtain an appropiate description it is necessary to quantize in a compatible way both gravity and matter fields. In particular, for a real--valued scalar field such quantization was developed in \cite{Ashtekar3}, and it is known as polymer quantization. This polymer representation is obtained by enforcing, at the quantum mechanical level, the diffeomorphism covariance through the unitary implementation of diffeomorphisms by applying non--regular representation techniques. As a consequence, the non--regularity conditions result in a representation which is not equivalent to the standard Schr\"odinger quantization, and when applied to minisuperspace models leads to a polymer--type representation known as loop quantum cosmology (LQC). Under this approach, several advances in the quantum gravity framework have been achieved, including microscopic states for black hole entropy \cite{Rovelli1},\cite{Ashtekar-Baez},\cite{Domagala}, the avoidance of classical singularities by quantum bounces \cite{Bojowald1},\cite{Bojowald2}, and the analysis of inhomogeneus perturbations in the cosmic expansion \cite{Agullo} (and references therein). Nonetheless, despite such significant efforts, fresh insights are needed in order to address major physical problems such as, a proper recovery of Einstein dynamics in the semiclassical limit, the fate of singularities in the full LQG and not only on symmetry reduced models, as well as a more suitable comprehension of the dynamics resulting from diffeomorphism invariant states \cite{Ashtekar4}.

In order to shed some light on these important issues, we propose to analyze the LQG program within the deformation quantization formalism. Deformation quantization, also refered as phase space quantum mechanics by many authors, consists of an alternative quantization procedure based on the idea that a quantum system is obtained by deforming the algebraic and geometrical structures of the classical phase space \cite{Bayen1},\cite{Bayen2}. One crucial element within this formulation lies on the definition of the Wigner function. This function corresponds to a quasi--probability distribution and is given by the phase space representation of the density matrix, which encodes the entire information about autocorrelation properties, expectation values and transition amplitudes. In other words, the Wigner function consists on the closest thing we have to a probability distribution in the quantum phase space. Despite  the deformation quantization formalism undoubtedly have provided important contributions to several areas in pure mathematics,  and also it has proved to be a reliable technique in the understanding of many physical quantum systems \cite{Waldmann}, the application of the formalism to contemporary problems of quantum gravity is poorly developed. Thus, one motivation of our work is to try to fill this gap.

Following some ideas developed in \cite{Corichi},\cite{DQpolymer}, the aim of this paper is to construct the Wigner functional corresponding to the polymer representation of the real--valued massive scalar field. As we will demostrate below, the polymer Wigner distribution is obtained as a limit, in the functional picture, of the Schr\"odinger representation with a Gaussian quantum weighted measure. The results presented here, to the best of our knowledge, are new and could pave the way in order to address key physical problems in LQG from a different perspective, since, according to recent developments on the perturbative formulations of algebraic quantum field theory, it has become clear that the star product and Wigner functions provide a valuable tool to formulate quantum field theories (QFT) in the semiclassical regime. Thereby, this current work corresponds to an ongoing progress. 

The paper is organized as follows, in section \ref{sec:Wpolymer} we review the Wigner function associated to the polymer representation of quantum mechanics. In section \ref{sec:Wignerscalar} we derive the Wigner--Weyl quantization scheme for the real massive scalar field within a quantum Gaussian weighted measure, and then define the corresponding Wigner functional. In section \ref{sec:Wignerscalarpoly} the polymer Wigner distribution is obtained as a functional limit measure. Finally, we introduce some concluding remarks in section \ref{sec:conclu}.

\section{The Wigner function of polymer quantum mechanics}
\label{sec:Wpolymer}
In this section we briefly review the polymer representation of quantum mechanics as a limit of the Schr\"odinger representation for the Weyl algebra in a Gaussian weighted measure. We will closely follow the description of the formalism as described in \cite{DQpolymer}. For simplicity, we focus on systems with one degree of freedom, nevertheless a generalization to more dimensions follows straightforwardly.

\subsection{The Wigner--Weyl quantization}
The simplest approach to quantize a classical system is to provide a quantization map, i.e. a one to one mapping $\mathcal{Q}_{\hbar}:\mathcal{A}\rightarrow \mathfrak{U}$ from the set of classical observables $\mathcal{A}=C^{\infty}(\mathbb{R}^{2})$, to the set of quantum observables $\mathfrak{U}$, usually given by self--adjoint operators defined on a Hilbert space $\mathcal{H}$. This map $\mathcal{Q}_{\hbar}$, depends on a positive parameter $\hbar>0$ and satisfies the properties
\begin{equation}
\lim_{\hbar\to 0}\frac{1}{2}\mathcal{Q}^{-1}_{\hbar}\left(\mathcal{Q}_{\hbar}(f_{1})\mathcal{Q}_{\hbar}(f_{2})+\mathcal{Q}_{\hbar}(f_{2})\mathcal{Q}_{\hbar}(f_{1}) \right)=f_{1}f_{2}, 
\end{equation}   
and
\begin{equation}
\lim_{\hbar \to 0}\mathcal{Q}_{\hbar}^{-1}\left(\frac{i}{\hbar}\left[\mathcal{Q}_{\hbar}(f_{1}),\mathcal{Q}_{\hbar}(f_{2}) \right]\right)=\left\lbrace f_{1},f_{2} \right\rbrace,  
\end{equation}
where $f_{1},f_{2}\in C^{\infty}(\mathbb{R}^{2})$. In general, it is well known that the relation defined by $\mathcal{Q}_{\hbar}$ between classical and quantum observables does not correspond to an isomorphism of Lie algebras, since in most of the cases, quantum phenomena differs from its classical counterpart. The mapping given by $\mathcal{Q}_{\hbar}$ occurs to be an isomorphism only in the limit $\hbar\to 0$, which according to the correspondence principle, quantum properties become classical \cite{Takhtajan}. Since quantum mechanics corresponds to a more accurate description of physical phenomena than classical mechanics, it is expected that the quantization process may not be unique.   

In our particular case, the quantization mapping $\mathcal{Q}_{\hbar}:\mathcal{A}\rightarrow\mathfrak{U}$, applied to a classical system defined on the phase space $\mathbb{R}^{2}$, with local coordinates $p$, $q$, stands for the passage from the Poisson bracket 
\begin{equation}
\left\lbrace q,p \right\rbrace=1, 
\end{equation}  
to the commutator of the operators
\begin{equation}\label{commutator}
\left[ \hat{Q},\hat{P}\right]\psi=i\hbar\psi,  
\end{equation}
where $\psi\in D$, a dense subset of the Hilbert space $\mathcal{H}$, such that the operators $\hat{Q}$, $\hat{P}$ and its commutator are defined. The equation (\ref{commutator}) is known as the canonical commutation relation. The quantization prescription for the position and momentum $\mathcal{Q}_{\hbar}(q)=\hat{Q}$, $\mathcal{Q}_{\hbar}(p)=\hat{P}$, satistying (\ref{commutator}), provides the standard cornerstone for the quantization of most of classical systems. 
In order to study the quantum kinematics on a particular Hilbert space $\mathcal{H}$, it is convenient to consider the algebra generated by the exponentiated versions of the operators $\hat{Q}$ and $\hat{P}$, expressed as
\begin{equation}
\hat{U}(u)=e^{-iu\hat{P}/\hbar}, \;\;\; \hat{V}(v)=e^{-iv\hat{Q}/\hbar},
\end{equation} 
where $u$ and $v$ denote real parameters with dimensions of length and momentum, respectively.
The canonical commutation relation in terms of the operators $\hat{U}(u)$ and $\hat{V}(v)$ reads
\begin{equation}
\hat{U}(u)\hat{V}(v)=e^{iuv/\hbar}\hat{V}(v)\hat{U}(u).
\end{equation}
The Weyl algebra, denoted by $\mathcal{W}$, is the algebra generated by taking finite linear combinations of the operators $\hat{U}(u)$ and $\hat{V}(v)$,
\begin{equation}
\sum_{i}\left( a_{i}\hat{U}(u_{i})+b_{i}\hat{V}(v_{i})\right)\in\mathcal{W}, \;\;\ \textrm{where}\;\; a_{i},b_{i}\in\mathbb{C}. 
\end{equation} 
From this perspective, defining a quantization mapping means finding a unitary representation of the Weyl algebra $\mathcal{W}$ on an arbitrary Hilbert space.

In order to obtain the standard Schr\"odinger representation one selects the Hilbert space, $\mathcal{H}_{Schr}=L^{2}(\mathbb{R},dq)$, the space of square--integrable functions with a translation invariant Lebesgue measure $dq$ on $\mathbb{R}$. Nevertheless, this representation is far from being unique. For our purposes, we instead choose to work with the Hilbert space
\begin{equation}
\mathcal{H}_{d}=L^{2}(\mathbb{R},d\mu_{d}),
\end{equation}  
given by the space of square integrable functions on $\mathbb{R}$ with respect to the Gaussian measure
\begin{equation}
d\mu_{d}=\frac{1}{d\sqrt{\pi}}e^{-\frac{q^{2}}{d^{2}}}dq,
\end{equation}
where $d$ is a parameter with dimensions of length. In the Hilbert space $\mathcal{H}_{d}$, the representation of the position and momentum operators take the form
\begin{equation}
\hat{Q}\psi(q)=(q\psi)(q)\;\;\;\textrm{and}\;\;\; \hat{P}\psi(q)=-i\hbar\frac{\partial}{\partial q}\psi(q)+i\hbar\frac{q}{d^{2}}\psi(q). 
\end{equation}
The unusual extra term acquired by the momentum operator is required in order to render it a self adjoint operator in the Gaussian measure. The main reason to select this particular representation stems from the Gelfand--Neimark--Segal (GNS) construction \cite{Strocchi}. Within this approach one determines the Hilbert space via the algebraic properties of the Weyl algebra $\mathcal{W}$, resulting in a representation that can be called of Fock--type, since creation and annihilation operators can be defined in a natural way.\\  
The different representations of the operator algebra turn out to be completely equivalent by means of the Stone--von Neumann uniqueness theorem, which asserts that all regular, irreducible representations of the Weyl algebra are unitarily equivalent. This imply that we can recover the Schr\"odinger representation from $\mathcal{H}_{d}$  through an isometric isomorphism $T:\mathcal{H}_{d}\rightarrow\mathcal{H}_{Schr}$ given by
\begin{equation}\label{isometry}
\psi(q):=T\phi(q)=\frac{1}{d^{1/2}\pi^{1/4}}e^{-\frac{q^{2}}{2d^{2}}}\phi(q),
\end{equation}
where $\psi\in\mathcal{H}_{Schr}$ and $\phi\in\mathcal{H}_{d}$. In this sense, all $d$--representations ($0<d<\infty$) in $\mathcal{H}_{d}$ are unitarely equivalent to the standard Schr\"odinger representation.

We turn now to the definition of a quantization prescription on $\mathcal{H}_{d}$. Following \cite{DQpolymer}, there is a linear map $\Phi$ from the set of classical observables given by $\mathcal{S}(\mathbb{R}^{2})$, the Schwartz space of functions defined on the phase space $\mathbb{R}^{2}$ whose derivatives are rapidly decreasing, into the linear operator space $\mathcal{L}(\mathcal{H}_{d})$. This map, called the Weyl quantization, is given by the formula
\begin{equation}
\Phi(f)\phi(q)=\frac{1}{2\pi\hbar}\int_{\mathbb{R}^{2}}f\left( p,\frac{q+q'}{2}\right)e^{\frac{i}{\hbar}p(q-q')}e^{-\frac{1}{2d^{2}}(q^{2}-q'^{2})}\phi(q')dpdq', 
\end{equation}
for $\phi\in\mathcal{H}_{d}$. The operator--valued mapping $\Phi$ is a Hilbert--Schmidt operator acting on $\mathcal{H}_{d}$, thus, it has a well defined trace but possibly infinite \cite{Reed}. Besides, the map $\Phi$ defines an homomorphism between $\mathcal{S}(\mathbb{R}^{2})$ and $\mathcal{L}(\mathcal{H}_{d})$, known as the Moyal product 
\begin{equation}
\Phi(f_{1})\Phi(f_{2})=\Phi(f_{1}* f_{2}),
\end{equation}
where the star product is given by
\begin{equation}\label{star}
(f_{1}*f_{2})(p,q)=f_{1}(p,q)\exp\left[-\frac{i\hbar}{2}\left(\overleftarrow{\partial}_{q}\overrightarrow{\partial}_{p}-\overleftarrow{\partial}_{p}\overrightarrow{\partial}_{q}\right)\right]f_{2}(p,q). 
\end{equation}

The inverse map $\Phi^{-1}$ associated to the Weyl quantizer $\Phi$, also known as the Weyl's inversion map, can be obtained as
\begin{equation}\label{Winv}
f(p,q)=\mathfrak{F}\left( \Tr(\Phi(f))e^{-\frac{iuv}{2\hbar}}V(-v)U(-u)\right), 
\end{equation}
where $f\in\mathcal{S}(\mathbb{R}^{2})$, $\mathfrak{F}$ stands for the Fourier transform operator $\mathfrak{F}:\mathcal{S}(\mathbb{R}^{2})\rightarrow\mathcal{S}(\mathbb{R}^{2})$
\begin{equation}
\tilde{f}(p,q)=\mathfrak{F}(f)(u,v)=\frac{1}{2\pi\hbar}\int_{\mathbb{R}^{2}}f(u,v)e^{-\frac{i}{\hbar}(up+vq)}dpdq,
\end{equation}
and the trace is taken with respect to any orthonormal basis for $\mathcal{H}_{d}$ \cite{Compean}.
With the Weyl quantization map and its inverse at hand, it is possible to obtain the Wigner function associated to the Hilbert space $\mathcal{H}_{d}$, which indeed, corresponds to the phase space representation of a quantum state. Let $\hat{\rho}$ denotes a density operator associated to a quantum state $\phi\in\mathcal{H}_{d}$, that is, a trace--one, self--adjoint and positive semi--definite operator written as
\begin{equation}\label{rho}
\hat{\rho}\phi(q)=\phi(q)\int_{\mathbb{R}}\overline{\varphi(q')}\chi(q')d\mu_{d}(q'),
\end{equation} 
(or $\hat{\rho}=\ket{\phi}\bra{\phi}$ in Dirac notation), where $\phi,\chi\in\mathcal{H}_{d}$. Since the density operator $\hat{\rho}$ corresponds to an integral operator (\ref{rho}), by means of the Weyl's inversion formula (\ref{Winv}) (see \cite{DQpolymer} for details), the phase space representation function associated to a quantum state is given by 
\begin{equation}\label{Wignerd}
\rho(\phi)(p,q)=\int_{\mathbb{R}}\phi\left(q+\frac{z}{2} \right)\overline{\phi\left(q-\frac{z}{2} \right) }e^{-\frac{i}{\hbar}zp}e^{-\frac{1}{d^{2}}(q^{2}+z^{2}/4)}\frac{dz}{d\sqrt{\pi}}. 
\end{equation}
This is the Wigner function defined on $\mathcal{H}_{d}$, a quasi--probability distribution in phase space, which is also normalized $\frac{1}{2\pi\hbar}\int_{\mathbb{R}^{2}}\rho(\phi)(p,q)dpdq=1$. In addition, the projections on the momentum and position leads to marginal probability densities \cite{DQpolymer}, usually called shadows
\begin{equation}
\mkern-50mu\frac{1}{2\pi\hbar}\int_{\mathbb{R}}\rho(\phi)(p,q)dp=||\phi||_{\mathcal{H}_{d}}^{2}, \;\;\; \textrm{and}\;\;\; \frac{1}{2\pi\hbar}\int_{\mathbb{R}}\rho(\phi)(p,q)dq=||\mathfrak{F}(T\phi)||_{\mathcal{H}_{Schr}}^{2}.
\end{equation} 

A counter--intuitive aspect of the Wigner function lies on the possibility to acquire negative values on certain regions over the phase space. Nevertheless, this odd feature makes the Wigner distribution so useful, since it allows us to visualize quantum trajectories in phase space and the negative probability values characterize joint--correlation functions and entanglement properties within the quantum system. In some sense, these imply that the Wigner function is the closest thing we have to a probability distribution in the quantum phase space \cite{Zachos}.\\ 
Finally, the Wigner function can be used to obtain the expectation value of an arbitrary operator $\hat{A}\in\mathcal{L}(\mathcal{H}_{d})$ as a phase space average 
\begin{equation}
\braket{\phi,\hat{A}\phi}_{\mathcal{H}_{d}}=\frac{1}{2\pi\hbar}\int_{\mathbb{R}^{2}}\rho(\phi)(p,q)A(p,q)dpdq, 
\end{equation}
where the operator $\hat{A}=\Phi(A)$ corresponds to the Weyl quantization transform of the classical function $A(p,q)$. In this sense, expectation values of quantum physical observables are computed through integration with respect to the Wigner distribution, in complete analogy with classical probability theory.

\subsection{The polymer representation}

Following \cite{Corichi},\cite{DQpolymer}, our purpose now is to obtain the polymer representation as a limit of the Wigner function defined on $\mathcal{H}_{d}$. The main idea is to study two limits of the Wigner distribution (\ref{Wignerd}) for the parameter $d$, the limit $1/d\to 0$, and the limit $d\to 0$. Both limits are well defined and the resulting representation agrees with the Wigner function associated with Loop Quantum Cosmology \cite{Sahlmann}. It is convenient to focus on the fundamental states in the Hilbert space $\mathcal{H}_{d}$, that is, those states generated by acting with $\hat{U}(u)$ and $\hat{V}(v)$ on the vacuum state $\phi_{0}=1$. Let us denote them by
\begin{equation}
\phi_{u}(q):=\hat{U}(u)\phi_{0}(q)=\Phi(U(u))\phi_{0}(q)=e^{\frac{u}{d^{2}}(q-\frac{u}{2})},
\end{equation} 
and
\begin{equation}\label{varphiv}
\phi_{v}(q):=\hat{V}(v)\phi_{0}(q)=\Phi(V(v))\phi_{0}(q)=e^{-\frac{i}{\hbar}vq},
\end{equation}
respectively. The derived states allow us to construct the corresponding Wigner function associated to polymer representation as a limiting case \cite{DQpolymer}. Starting with the Wigner function defined in (\ref{Wignerd}) for the vector states $\varphi_{v}$, the limit $1/d\to 0$ gives us 
\begin{equation}\label{polyA}
\lim_{1/d\to 0}\rho(\phi_{v})(p,q)=\delta_{p,-v}=:\rho_{A}(p,q),
\end{equation}
similarly, for the limiting case $d\to 0$ associated to the Wigner function defined by the states $\phi_{u}$, we obtain
\begin{equation}\label{polyB}
\lim_{d\to 0}\rho(\phi_{u})(p,q)=\delta_{q,u}=:\rho_{B}(p,q).
\end{equation}
Both limits correspond to the A and B--polymer representations of quantum mechanics \cite{Corichi}, and as we can observe from expresions (\ref{polyA}) and (\ref{polyB}), the obtained Wigner distributions distinctly reflect the way in which the wave functions are modulated by Kronecker deltas on a countable (or possibly uncountable) number of points. To be more precise, within the polymer representation the Hilbert space consists of functions that vanishes everywhere except for a countable number of points, which usually are taken to form a regularly spaced lattice $q_{n}=q_{0}+n\lambda$, for a given $q_{0}\in\mathbb{R}$ and $n\in\mathbb{Z}$. For a fixed point $q_{0}$ , the wave functions, supported on this lattice, actually belong to a separable Hilbert space which corresponds to a superselected sector of the full polymer Hilbert space, i.e., $\mathcal{H}_{poly}=\oplus_{q_{0}\in[0,\lambda]}\mathcal{H}_{q_{0}}$. Furthermore, $\rho_{A}$ and $\rho_{B}$ prove to be equivalent to the Wigner functions associated to the pure characters and their Fourier transforms over the Bohr compactification  of the real line $\mathbb{R}_{B}$, that is
\begin{equation}\label{WignerLQCA}
\rho_{A}(p,q)=\int_{\mathbb{R}_{B}}\phi_{v}\left(q+\frac{1}{2}b \right)\overline{\phi_{v}\left( q-\frac{1}{2}b\right)}h(b,-p)db, 
\end{equation} 
\begin{equation}\label{WignerLQCB}
\rho_{B}(p,q)=\int_{\hat{\mathbb{R}}_{B}}\tilde{\phi}_{u}\left(q+\frac{1}{2}\tau \right)\overline{\tilde{\phi}_{u}\left( q-\frac{1}{2}\tau\right)}h(b,\tau)d\tau,
\end{equation}
where $\mathbb{R}_{B}$ denotes the Bohr compactification of the reals, $\hat{\mathbb{R}}_{B}$ stands for its locally compact dual group and $h(b,p)=e^{ipb}$ are the characters of $\mathbb{R}_{B}$. This means, that the right hand side of equations (\ref{WignerLQCA}), (\ref{WignerLQCB}) are precisely the Wigner function for Loop Quantum Cosmology (LQC), in the position and momentum representation respectively, and they were obtained by using the fundamental vectors states $\phi_{v}$ and $\phi_{u}$ generated by the Weyl algebra \cite{Sahlmann}. Since, in the context of LQC, a generic wave function is given by a finite span of the functions $\phi_{v}$, namely the set of cylindrical functions $Cyl(\mathbb{R}_{B})$, the linearity properties of the Wigner function $\rho(p,q)$ in $\mathcal{H}_{d}$ (\ref{Wignerd}), implies that the Wigner distribution for LQC (or polymer Wigner distribution) can be obtained through a limiting process of the standard Wigner function defined on a Gaussian weighted measure space.

In the following section, we will implement the polymer representation of quantum mechanics in order to obtain the polymer Wigner distribution for a scalar field. 

\section{The Wigner function for the scalar field}
\label{sec:Wignerscalar}
In this part, we shall derive the Wigner--Weyl quantization scheme for the massive scalar field in a Gaussian measure defined on an infinite dimensional vector space.

Consider a real scalar field $\varphi$ defined on a 4--dimensional background Minkowski spacetime $\mathcal{M}$. Let us perform a 3+1 decomposition of the spacetime in the form $\mathcal{M}=\Sigma\times\mathbb{R}$, for any  Cauchy surface $\Sigma$, which in the present case is topologically equivalent to $\mathbb{R}^{3}$. The spacetime manifold $\mathcal{M}$ is endowed with a metric $\eta=diag(+1,+1,+1,-1)$ and local coordinates $(x,t)\in\mathbb{R}^{3}\times\mathbb{R}$. For simplicity, we deal with fields at the instant $t=0$ and write $\varphi(x,0):=\varphi(x)$, and $\pi(x,0):=\pi(x)$, where $\pi(x)$ stands for the canonical conjugate momentum associated to $\varphi(x)$. Thus, the phase space of the theory is locally written as $\Gamma=(\varphi,\pi)$ and can be related to as suitable initial data associated to a Cauchy surface $\Sigma$.

According to \cite{DQpolymer} and \cite{Compean}, in order to construct the Wigner function for the scalar field, we need to provide a quantization map such that it takes the Poisson bracket 
\begin{equation}
\left\lbrace \varphi(x),\pi(y) \right\rbrace=\delta(x-y), 
\end{equation}
to the commutator of operators
\begin{eqnarray}\label{CCR}
\left[\hat{\varphi}(x),\hat{\pi}(y) \right]\Psi &=& i\hbar\delta(x-y)\Psi,\nonumber \\
\left[\hat{\varphi}(x),\hat{\varphi}(y) \right]\Psi &=& \left[\hat{\pi}(x),\hat{\pi}(y) \right]\Psi=0,  
\end{eqnarray}
where the state $\Psi[\varphi]$, at least in an intuitive level, is given by a functional of the field $\varphi \in \mathcal{S}'(\mathbb{R}^{3})$. To be more specific, $\Psi$ belongs to the Hilbert space $\mathcal{H}_{S}=L^{2}(\mathcal{S}'(\mathbb{R}^{3}),d\mu)$, where $\mathcal{S}'(\mathbb{R}^{3})$ represents the Schwartz space of tempered distributions, that is, the space of continuous linear functionals on the Schwartz space of rapidly decreasing smooth test functions $\mathcal{S}(\mathbb{R}^{3})$, and $d\mu$ is a measure to be specified \cite{Glimm}. Analogoulsy to section (\ref{sec:Wpolymer}), implementing a quantization process means to find a representation of the Weyl algebra $\mathcal{W}$, generated by the finite linear combination of the exponential version of the operators $\hat{\pi}$ and $\hat{\varphi}$, defined as
\begin{equation}\label{Walgebra}
\hat{U}(u)=e^{-\frac{i}{\hbar}\int_{\mathbb{R}^{3}}dx\,u\hat{\pi}}, \;\;\; \hat{V}(v)=e^{-\frac{i}{\hbar}\int_{\mathbb{R}^{3}}dx\,v\hat{\varphi}},
\end{equation}
for $u$, $v\in\mathcal{S}(\mathbb{R}^{3})$. Whenever $\hbar\neq 0$, the operators $\hat{U}(u)$ and $\hat{V}(v)$ satisfy the commutation relation
\begin{equation}
\hat{U}(u)\hat{V}(v)=e^{-\frac{i}{\hbar}\braket{u,v}}\hat{V}(v)\hat{U}(u),
\end{equation}
where $\braket{u,v}$ denotes the inner product on $L^{2}(\mathbb{R}^{3})$, that is
\begin{equation}
\braket{u,v}=\int_{\mathbb{R}^{3}}dx\,uv.
\end{equation}
From this point of view, determining a quantization mapping means to provide a unitary representation of the Weyl algebra $\mathcal{W}$ on the Hilbert space $\mathcal{H}_{S}$.

In order to construct the Schr\"odinger representation of the Weyl algebra $\mathcal{W}$, let us consider a Gaussian measure $d\mu_{C}$, of mean zero and covariance $C:=(-\Delta+m^{2})^{-1/2}$. It existence relies on the Bochner--Minlo's theorem \cite{Reed}, \cite{Gelfand}, which in our case asserts that given $C$ a linear, positive and self--adjoint operator on $\mathcal{S}(\mathbb{R}^{3})$, then there exists a unique normalized measure $d\mu_{C}$, of mean zero, such that for all $f\in\mathcal{S}(\mathbb{R}^{3})$ satisfies
\begin{equation}\label{GM}
\chi(f):=e^{-\braket{f,Cf}}=\int d\mu_{C}(\varphi)\,e^{i\varphi(f)},
\end{equation}  
where we have used the notation
\begin{equation}
\varphi(f)=\int_{\mathbb{R}^{3}} dx\,\varphi f.
\end{equation} 
The term $\chi(f)$, defined through equation (\ref{GM}), represents the Fourier transformation of the measure $d\mu_{C}$ evaluated at $f$. Furthermore, $\chi\in\mathcal{S}'(\mathbb{R}^{3})$ and is also called the generating or characteristic functional, since from it, we can compute the moments of the measure $d\mu_{C}$ as
\begin{equation}
\int d\mu_{C}\,\varphi(f)^{n}=\left.\left(-i\frac{d}{d\lambda} \right)^{n}\chi(\lambda f)\right|_{\lambda=0}.
\end{equation}
Explicitly, the Gaussian measure $d\mu_{C}$ corresponds to a measure of the form
\begin{equation}\label{GMe}
d\mu_{C}=e^{-\braket{\varphi,C^{-1}\varphi}}\mathcal{D}\varphi=e^{-\int_{\mathbb{R}^{3}}dx\,\varphi(-\Delta+m^{2})^{1/2}\varphi}\mathcal{D}\varphi,
\end{equation}
where the formal expression of  $\mathcal{D}\varphi=\prod_{x\in\mathbb{R}^{3}}d\varphi(x)$, denotes something like a uniform Lebesgue measure on the configuration space. However, it is known that in an infinite dimensional vector space a translational invariant measure cannot be properly defined. This implies that the form of the measure $d\mu_{C}$ in equation (\ref{GMe}), although is not completely well defined, contrary to the expression $\chi(f)$ in (\ref{GM}), it proves to be useful for understanding the denomination of Gaussian measure. This can be seen by approximating the measure by using Riemann sums. Let us consider a $3$-dimensional cubic lattice with lattice spacing $a$. The integral in the exponential term of the measure (\ref{GMe}) is approximated by
\begin{equation}
\frac{1}{a^{3}}\sum_{i,j\in\mathbb{Z}^{3}}\braket{\varphi(ai),(-\Delta+m^{2})^{1/2}_{ij}\varphi(aj) }, 
\end{equation}
where $(-\Delta+m^{2})^{1/2}_{ij}$ denotes the finite difference approximation of the inverse of the covariance operator $C$. Now, by summing over all finite volumes $\Lambda$, we can consider the measure
\begin{equation}
\mkern-105mu d{\mu}_{C}(a,\Lambda)=\frac{1}{N(a,\Lambda)}\exp\left(-\frac{1}{a^{3}}\sum_{i,j\in\mathbb{Z}^{3}\cap\Lambda} \braket{\varphi(ai),(-\Delta+m^{2})^{1/2}_{ij}\varphi(aj) }\right)\prod_{k\in\mathbb{Z}^{3}\cap \Lambda}d\varphi(ak), 
\end{equation}
where $N(a,\Lambda)$ is a constant such that the integral is normalized to one. We observe that the integration upon this restricted measure over a lattice gives precisely the term associated to the left hand side of equation (\ref{GM}), moreover
\begin{equation}
\lim_{a\to 0}\lim_{\Lambda\to\mathbb{R}^{3}}d\mu_{C}(a,\Lambda)=d\mu_{C},
\end{equation}
assuming that the finite difference operator converges to $C^{-1}$ \cite{Glimm}. 

In standard quantum mechanics the choice of the measure, somehow, turns out to be arbitrary. The reason lies on the Stone--von Neumann theorem, which assures us that any representation of the Weyl algebra is unitarely equivalent to the Schr\"odinger representation. In case of field theories this is no longer true, and in fact, there are infinitely many inequivalent representations of the Weyl algebra $\mathcal{W}$.

The motive to select this Gaussian measure is twofold. First, from the algebraic Quantum Field Theory perspective, the main idea is to formulate the quantum theory for a real Klein--Gordon field by considering the observables as the relevant objects, and relegate the states as secondary entities that act on the observables. In order to obtain a representation of the Weyl algebra $\mathcal{W}$, and since there are infinitely many of them, one makes use of the Fock representation supplied by the canonical commutation relations. Within this representation, a complex structure on the phase space must be specified (compatible with Lorentz invariance) and then, the expectation values of the Weyl operators are obtained via the Fock vacuum. Ultimately, the computed expectation values define a positive linear functional $\omega_{fock}$ on the algebra of observables. This is precisely the moment where the Gaussian measure comes into play. For the Schr\"odinger representation will be equivalent to the Fock formulation, the positive linear functional $\omega_{fock}$, must be equal to the algebraic state obtained via the GNS construction. This state is defined through a measure which proves to be the Gaussian measure defined in (\ref{GM}) \cite{Fock}, \cite{Corichi2}.\\
The second reason to favour the Gaussian measure $d\mu_{C}$ over the Lebesgue--like measure $\mathcal{D}\varphi$, is simply because the measure $d\mu_{C}$ corresponds to a proper probability measure, defined by means of the expression (\ref{GM}), while the formal uniform measure $\mathcal{D}\varphi$ is not completely well--defined. Moreover, it can be shown that the Gaussian measure provides, for each value of the mass, a consistent quantization of the dynamics with a unique symmetry--invariant vacuum state \cite{Baez}, \cite{Velinho}. 

In the Hilbert space $\mathcal{H}_{S}=L^{2}(\mathcal{S}'(\mathbb{R}^{3}),d\mu_{C})$, the abstract field operators $\hat{\varphi}$ and $\hat{\pi}$ are represented as
\begin{equation}\label{position}
\hat{\varphi}\Psi[\varphi]=\varphi\Psi[\varphi],
\end{equation}
and
\begin{equation}\label{momentum}
\hat{\pi}\Psi[\varphi]=-i\hbar\frac{\delta}{\delta\varphi}\Psi[\varphi]+i\hbar(-\Delta+m^{2})^{1/2}\varphi\Psi[\varphi], 
\end{equation}       
where $\delta/\delta\varphi$ denotes the functional derivative, and the second term in (\ref{momentum}) appears in order to render the momentum operator self--adjoint with respect to the Gaussian mesaure $d\mu_{C}$.\\
Next, in order to find the quantization mapping, we define $\hat{S}(u,v)\in\mathcal{L}(\mathcal{H}_{S})$ as a linear operator on $\mathcal{H}_{S}$ given by
\begin{equation}\label{S}
\hat{\mathcal{S}}(u,v):=e^{-\frac{i}{2\hbar}\braket{u,v}}\hat{U}(u)\hat{V}(v).
\end{equation}
From the canonical commutation relations (\ref{CCR}), and the representations of the field operators (\ref{position}), (\ref{momentum}), this operator follows the identities
\begin{equation}
\hat{S}(u_{1},v_{1})\hat{S}(u_{2},v_{2})=e^{-\frac{i}{2\hbar}(\braket{u_{1},v_{2}}-\braket{u_{2},v_{1}})}\hat{S}(u_{1}+u_{2},v_{1}+v_{2}),
\end{equation}
\begin{equation}
\hat{S}(u,v)^{\dagger}=\hat{S}(-u,-v),
\end{equation}
and
\begin{equation}\label{traceS}
\tr{(\hat{S}(u,v)\hat{S}(u',v'))}=\delta(u-u')\delta(v-v'),
\end{equation}
where $\hat{S}(u,v)^{\dagger}$ denotes the adjoint operator associated to $\hat{S}(u,v)$, and the trace is taken with respect to an orthonormal basis for $\mathcal{H}_{S}$ \cite{Compean}. Now, let us define the linear map $\Phi:L^{2}(\Gamma)\to \mathcal{L}(\mathcal{H}_{S})$, from the space of functionals on the phase space $\Gamma$ to the linear operatos acting on the Hilbert space $\mathcal{H}_{S}$ by (in the following we will take $\hbar$=1),
\begin{equation}
\Phi(F)\Psi[\varphi]=\int\mathcal{D}u\mathcal{D}v\,\tilde{F}(u,v)\hat{S}(u,v)\Psi[\varphi],
\end{equation}
where $\tilde{F}(u,v)$ stands for the inverse Fourier transform of the functional $F\in L^{2}(\Gamma)$, that is
\begin{equation}
\tilde{F}(u,v)=\int\mathcal{D}\pi\mathcal{D}\varphi\,e^{-i(\braket{u,\pi}+\braket{v,\varphi})}F(\pi,\varphi).
\end{equation}
This map $\Phi$ is called the Stratonovich--Weyl quantizer \cite{Compean}, and should be understood as a map from the functionals defined on the classical phase space to linear operators, such that
\begin{equation}
\braket{\Phi(F)\Psi_{1},\Psi_{2}}_{\mathcal{H}_{S}}=\int\mathcal{D}u\mathcal{D}v\,\tilde{F}(u,v)\braket{\hat{S}(u,v)\Psi_{1},\Psi_{2}}_{\mathcal{H}_{S}},
\end{equation}   
is absolutely convergent and $\braket{\cdot,\cdot}_{\mathcal{H}_{S}}$ denotes the inner product defined by the Gaussian measure $d\mu_{C}$.
By using the explicit expressions for $\hat{S}(u,v)$ in (\ref{S}) in terms of the operators (\ref{position}), (\ref{momentum}) and the inverse Fourier transform, the Stratonovich--Weyl quantization map reads
\begin{equation}\label{SWmap}
\!\!\!\!\!\!\!\!\!\!\!\!\!\Phi(F)\Psi[\varphi]=\int\mathcal{D}\pi\mathcal{D}\varphi'\, F\left[ \pi,\frac{\varphi+\varphi'}{2}\right]e^{i(\braket{\pi,\varphi-\varphi'}+\frac{1}{2}\braket{\varphi,C^{-1}\varphi}-\frac{1}{2}\braket{\varphi',C^{-1}\varphi'})}\Psi[\varphi'], 
\end{equation} 
for any $\Psi\in\mathcal{H}_{S}$. This means that the operator $\Phi$ represents a kind of an integral operator
\begin{equation}
\Phi(F)\Psi[\varphi]=\int\mathcal{D}\varphi'\,K(\varphi,\varphi')\Psi[\varphi'],
\end{equation}
where the Kernel $K(\varphi,\varphi')$ is given by
\begin{equation}
K(\varphi,\varphi')=\int\mathcal{D}\pi\, F\left[ \pi,\frac{\varphi+\varphi'}{2}\right]e^{i(\braket{\pi,\varphi-\varphi'}+\frac{1}{2}\braket{\varphi,C^{-1}\varphi}-\frac{1}{2}\braket{\varphi',C^{-1}\varphi'})}.
\end{equation}
The inverse map $\Phi^{-1}:\mathcal{L}(\mathcal{H}_{S})\to L^{2}(\Gamma)$ associated to the Stratonovich--Weyl quantizer, also known as the Weyl's inversion formula, can be obtained by multiplying (\ref{SWmap}) by $\hat{S}(\pi,\varphi)$ and taking the trace using property (\ref{traceS}), that is
\begin{equation}\label{Winversion}
F(\pi,\varphi)=\tr(\Phi(F)\hat{S}(\pi,\varphi)).
\end{equation}

Now, it is possible to define the Wigner function in the following manner. Let $\hat{\rho}$ be a density operator associated to a quantum state $\Psi\in\mathcal{H}_{S}$, that is, a positive semi--definite operator written as
\begin{equation}
\hat{\rho}\Psi'[\varphi]=\Psi[\varphi]\int d\mu_{C}(\varphi')\,\overline{\Psi[\varphi']}\Psi'[\varphi'],
\end{equation}
(or $\hat{\rho}=\ket{\Psi}\bra{\Psi}$ in Dirac notation), where $\Psi,\Psi'\in\mathcal{H}_{S}$. From the Weyl's inversion formula (\ref{Winversion}), the phase--space functional $\rho(\pi,\varphi)$ associated to the density operator $\hat{\rho}$ is given by
\begin{equation}\label{Wignerfield}
\rho(\pi,\varphi)=\int\mathcal{D}\varphi'\,\Psi[\varphi+\frac{\varphi'}{2}]\overline{\Psi[\varphi-\frac{\varphi'}{2}]}e^{-i\braket{\varphi',\pi}}e^{-\braket{\varphi,C^{-1}\varphi}-\frac{1}{4}\braket{\varphi',C^{-1}\varphi'}}.
\end{equation}
Thus, the functional $\rho$ represents the Wigner function for a real Klein--Gordon field with respect to a Gaussian quantum measure $d\mu_{C}$, with covariance $C=(-\Delta+m^{2})^{-1/2}$. It must be noted that, the expression for the Wigner functional obtained in (\ref{Wignerfield}), provides us with a generalization in the case of fields, of the Wigner function corresponding to systems with a finite number of degrees of freedom as calculated in (\ref{Wignerd}). 

Finally, to finish this section and in order to compare our developed formalism with other phase--space prescriptions where the Wigner function for the massive scalar field is also been obtained, let us calculate the Wigner functional for to the ground state. In the Schr\"odinger representation associated to the Hilbert space $\mathcal{H}_{S}$, with Gaussian measure $d\mu_{C}$, one can prove that the vacuum state $\Psi_{0}$ is given by the constant functional $\Psi_{0}[\varphi]=1\in\mathcal{H}_{S}$ \cite{Corichi}. Then, by substituting the state $\Psi_{0}$ on (\ref{Wignerfield}) and performing some integral calculations by using property (\ref{GM}), one finds that the Wigner functional $\rho_{0}$ of the ground state reads
\begin{equation}
\rho_{0}(\pi,\varphi)=e^{-\braket{\varphi,C^{-1}\varphi}-\braket{\pi,C\pi}},
\end{equation}  
which is exactly the Wigner functional associated to the vacuum state according to \cite{Compean}, \cite{ZachosField}. In particular, the quantum fields and the Wigner functional determined from the Gaussian measure (\ref{GMe}), can be readily applied to the case of the massless scalar field. If the dimension of the spacetime is equal to two, the Hilbert space can be constructed by choosing a different test function space where the covariance operator $C$ is defined. Let $\mathcal{S}_{0}(\mathbb{R})=\left\lbrace f\in\mathcal{S}(\mathbb{R}):\mathfrak{F}(f)(0)=0)\right\rbrace$ a subspace of the Schwartz space of rapidly decreasing test functions, then, the Bochner-Minlo's theorem gaurantees the existence of a measure in the dual space $\mathcal{S}'_{0}(\mathbb{R})$, which yields the zero mass case as showed in \cite{Glimm}. Moreover, if the dimension of the spacetime is greater than two, one needs to define a Gaussian measure supported on the space of generalized distributions $\mathcal{D}'(\mathbb{R}^{n})$, that is, the dual space associated to the smooth functions of compact support $\mathcal{D}(\mathbb{R}^{n})$. Once again, the Bochner-Minlo's theorem asserts that there is a unique Gaussian measure in the space $\mathcal{D}'(\mathbb{R}^{n})$, where the states and operators corresponding to the zero mass case are well defined \cite{Gelfand},\cite{Asorey}. In the next section we will see how the Wigner functional $\rho$ can be used to obtain the polymer representation of the scalar field.

\section{The polymer representation of the scalar field}
\label{sec:Wignerscalarpoly}

In this section we derive the polymer representation of the massive scalar field from the Schr\"odigner representation with a Gaussian measure $d\mu_{C}$, within the formalism of Wigner--Weyl quantization developed in the previous section. We will first analyze the Weyl algebra by means of the Stratonovich--Weyl quantizer, and then through some limiting process, we will obtain the Wigner functional associated to the polymer representation.

As a first step, we  need to stablish how the Weyl algebra $\mathcal{W}$, generated by the operators $\hat{U}(u)$ and $\hat{V}(v)$ defined on (\ref{Walgebra}), are represented on any $\Psi[\varphi]\in\mathcal{H}_{S}=L^{2}(\mathcal{S}'(\mathbb{R}^{3}),d\mu_{C})$. By employing the Stratonovich--Weyl map (\ref{SWmap}), we obtain
\begin{eqnarray}
\!\!\!\!\!\!\!\!\!\!\!\!\!\!\Phi(U(u))\Psi[\varphi]&=&\int\mathcal{D}\pi\mathcal{D}\varphi'\,U\left(\pi,\frac{\varphi+\varphi'}{2} \right)e^{i(\braket{\pi,\varphi-\varphi'}+\frac{1}{2}\braket{\varphi,C^{-1}\varphi}-\frac{1}{2}\braket{\varphi',C^{-1}\varphi'})}\Psi[\varphi'], \nonumber \\
&=&e^{\braket{u,C^{-1}\varphi}+\frac{1}{2}\braket{u,C^{-1}u}}\Psi[\varphi-u],
\end{eqnarray} 
and
\begin{eqnarray}
\!\!\!\!\!\!\!\!\!\!\!\!\!\!\Phi(V(v))\Psi[\varphi]&=&\int\mathcal{D}\pi\mathcal{D}\varphi'\,V\left(\pi,\frac{\varphi+\varphi'}{2} \right)e^{i(\braket{\pi,\varphi-\varphi'}+\frac{1}{2}\braket{\varphi,C^{-1}\varphi}-\frac{1}{2}\braket{\varphi',C^{-1}\varphi'})}\Psi[\varphi'], \nonumber \\
&=&e^{-i\braket{v,\varphi}}\Psi[\varphi].
\end{eqnarray} 
Before proceeding, and according to the algebraic formulation of QFT \cite{Corichi2}, it is convenient to focus on the fundamental vector states in the Hilbert space $\mathcal{H}_{S}$, that is, those vectors generated by the action of the Weyl algebra operators $\hat{U}(u)$ and $\hat{V}(v)$ on the vacuum state $\Psi_{0}=1$. Let us call them, for reasons that will be clear afterwards,
\begin{equation}
\Psi_{\varphi}[\varphi]:=\hat{U}(u)\Psi_{0}[\varphi]=\Phi(U(u))\Psi_{0}[\varphi]=e^{\braket{u,C^{-1}\varphi}-\frac{1}{2}\braket{u,C^{-1}u}},
\end{equation}
and
\begin{equation}\label{Cyl}
\Psi_{\pi}[\varphi]:=\hat{V}(v)\Psi_{0}[\varphi]=\Phi(V(v))\Psi_{0}[\varphi]=e^{-i\braket{v,\varphi}},
\end{equation}
respectively, where we have used that $\Psi_{0}[\varphi]=1$ in each of the last identities.

Now, let us analyze the Wigner functionals corresponding to the fundamental states $\Psi_{\varphi}$ and $\Psi_{\pi}$ defined above. Starting with the Wigner functional established in (\ref{Wignerfield}) for the vector states $\Psi_{\pi}$, we obtain
\begin{eqnarray}\label{Wignerpi}
\rho_{\pi}(\pi,\varphi)&=&\int\mathcal{D}\varphi'\,\Psi_{\pi}[\varphi+\frac{\varphi'}{2}]\overline{\Psi_{\pi}[\varphi-\frac{\varphi'}{2}]}e^{-i\braket{\varphi',\pi}}e^{-\braket{\varphi,C^{-1}\varphi}-\frac{1}{4}\braket{\varphi',C^{-1}\varphi'}}, \nonumber \\
&=&e^{-\braket{\varphi,C^{-1}\varphi}}e^{-\braket{v+\pi,C(v+\pi)}}.
\end{eqnarray}
We observe that if the inverse of the covariance operator $C^{-1}\to 0$ as a bilinear form in a weak operator convergence, that is, if $\braket{\xi,C^{-1}\xi}\to 0$ for all $\xi\in \mathcal{S}'(\mathbb{R}^{3})$, then $\rho_{\pi}$ converges to 
\begin{equation}\label{Wignerpipoly}
\rho_{\pi}(\pi,\varphi)\to\delta_{\pi,-v}=:\rho_{\pi}^{poly}(\pi,\varphi),
\end{equation}
since, within this limit, the exponentials converge to $e^{-\braket{\varphi,C^{-1}\varphi}}\to 1$ and $e^{-\braket{v+\pi,C(v+\pi)}}\to \delta_{\pi,-v}$, respectively. This expression corresponds to the Wigner functional associated to the "$\pi$--polarization" of the polymer representation of the scalar field, derived using algebraic states via the GNS construction \cite{Chung}.
Similarly, for the limiting case $C\to 0$ weakly, the Wigner functional defined by the vector states $\Psi_{\varphi}$, converges to
\begin{equation}\label{Wignervarphipoly}
\rho_{\varphi}(\pi,\varphi)\to\delta_{\varphi,u}=:\rho_{\varphi}^{poly}(\pi,\varphi),
\end{equation}
which is equivalent to the "$\varphi$--polarization" of the polymer representation described in \cite{Chung}. As can be seen, both limits are well defined and in fact correspond to the degenerate case of the limit of Gaussian measures through weak convergence of covariance operators, as illustrated on \cite{Glimm}.


In order to understand the meaning of the Wigner functionals $\rho_{\pi}^{poly}(\pi,\varphi)$ and $\rho_{\varphi}^{poly}(\pi,\varphi)$, associated to the polymer representations for the massive scalar field, let us express the Wigner functional (\ref{Wignerpi}), in terms of the Fourier decomposition of the fields $\varphi, \pi$ and $v$, we have
\begin{equation}\label{Wpik}
\rho_{\pi}=e^{-\int dk\,\omega_{k}\tilde{\varphi}_{k}^{2}}e^{-\int dk\,\frac{1}{\omega_{k}}(\tilde{v}_{k}+\tilde{\pi}_{k})^{2}}=\prod_{k}e^{-\omega_{k}\tilde{\varphi}^{2}_{k}}e^{-\frac{1}{\omega_{k}}(\tilde{v}_{k}+\tilde{\pi}_{k})^{2}},
\end{equation}     
where $\omega_{k}=\sqrt{k^{2}+m^{2}}$ and $\tilde{\varphi}_{k}$, $\tilde{v}_{k}$ and $\tilde{\pi}_{k}$ represent the $k$--modes of the fields. With the aim to recover the polymer representation, a formal limit of the frequencies $\omega_{k}$ can be taken as in \cite{Chung}. Nevertheless, in the present context, we note that the Wigner functional $\rho_{\pi}$ depicted in (\ref{Wpik}), is given by the infinite product of Wigner functions for the vector states $\phi_{v}\in\mathcal{H}_{d}$, defined in (\ref{varphiv}), where the parameter $d$ plays an analogous role to the frequency $\omega_{k}$ introduced here \cite{DQpolymer}. From (\ref{isometry}), we also know that all $d$--representations (with $d>0$) are unitarely equivalent. Therefore, for convenience, let us make the substitution $\omega_{k}\mapsto\omega_{k}/d^{2}$, with $d$ a real parameter. Then, the resulting Wigner functional associated with the limit $1/d\to 0$ reads
\begin{equation}\label{Wignerlimit}
\lim_{1/d\to 0}\rho_{\pi}=\lim_{1/d\to 0}\prod_{k}e^{-\frac{\omega_{k}}{d^{2}}\tilde{\varphi}^{2}_{k}}e^{-\frac{d^{2}}{\omega_{k}}(\tilde{v}_{k}+\tilde{\pi}_{k})^{2}}=\prod_{k}\delta_{\pi_{k},-v_{k}}=\prod_{k}\rho_{A_{k}}=\rho^{poly}_{\pi},
\end{equation}     
where $\rho_{A_{k}}$ stands for Wigner function associated to the A--polymer representation (\ref{polyA}) for each mode $k$.

The previous analysis shows that, the polymer Wigner functional $\rho_{\pi}^{poly}(\pi,\varphi)$ corresponding to the massive scalar field in the $\pi$--polarization (\ref{Wignerpipoly}), can be obtained by replacing, for each Fourier $k$--mode of the scalar field, the Wigner function associated to the A--polymer representation (\ref{polyA}). Then, we conclude that the weak limit of the covariance operator $C^{-1}\to 0$ in (\ref{Wignerpi}), corresponds to the replacement of each Fourier mode of the massive scalar field by its polymer analog. Similarly, for the case of the $\varphi$--polarization, is easy to verify that the role implemented by the A--polymer representation is now done by the B--polymer representation. These results prove to be consistent and in a complete agreement with the polymer representation of the real--valued scalar field concluded by other methods, such as the GNS construction and the Fock quantization \cite{Chung}, \cite{Husain}.

In the particular case of the massless scalar field, the Wigner functional associated polymer representation can also be derived in principle, by performing the procedure developed above, but now on the Hilbert space $\mathcal{H}_{S}=L^{2}(\mathcal{D}'(\mathbb{R}^{3}),d\mu_{C})$, where $d\mu_{C}$ corresponds to the Gaussian measure defined by the Bochner-Minlo's theorem in the space on the generalized distributions $\mathcal{D}'(\mathbb{R}^{3})$. However, since the support of the measures associated to the massive and massless cases differ, it is known that both measures are in fact singular with respect to each other. Therefore, in order to study possible differences ocurring in the convergence of the covariance operators, an analysis of the large scale behaviour of distributions is required \cite{measures}.         

Finally, let us now discuss some features of the polymer quantization developed within our formalism. In LQG, matter fields can only have support on polymer--like excitations of quantum geometry, this implies that the quantum states cannot refer to any classical or even continuous background geometry \cite{Ashtekar1},\cite{Ashtekar2},\cite{Rovelli}. In order to accomplish these requirements on a quantum geometry, one consider a set $V=\left\lbrace x_{1},\ldots, x_{n}\right\rbrace $, given by a finite set of points on $\mathbb{R}^{3}$ called a vertex set. Then, the $C^{\star}$--algebra of configuration observables for the scalar field are generated by the space of Cylindrical functions $\textrm{Cyl}:=\cup_{V}\textrm{Cyl}_{V}$, where $\textrm{Cyl}_{V}$ represents the cylindrical functions supported on a vertex set $V$, that is, finite linear combinations of the functions depending on the field $\varphi$
\begin{equation}
\mathcal{N}_{V,v}(\varphi)=e^{-i\sum_{j}v_{j}\varphi(x_{j})},
\end{equation}   
where $v_{1},\ldots,v_{n}$ are arbitrary real parameters. The functions $\mathcal{N}_{V,v}(\varphi)$, also called scalar network functions according with the terminology of LQG, can be obtained by using the fundamental states $\Psi_{\pi}$ defined on (\ref{Cyl}), and taking the function $v$ as a function supported on the vertex set $V$. This implies, that the polymer Wigner functional (\ref{Wignerpi}) for the scalar network states $\mathcal{N}_{V,v}(\varphi)$, is written as
\begin{equation}
\rho^{poly}_{\pi, V}(\pi,\varphi)=\prod_{x_{j}\in V}\delta_{\pi(x_{j}),-v_{j}}.
\end{equation}
Note that $\rho^{poly}_{\pi,V}(\pi,\varphi)$, corresponds to a finite product of the polymer Wigner functions in the A--representation (\ref{polyA}), and each of them is equivalent to a Wigner function defined over the Bohr compactification of the real line $\mathbb{R}_{B}$ (\ref{WignerLQCA}). This means, that the quantum configuration space of the polymer representation of the scalar field, given in terms of Wigner distributions, consists of $\mathbb{R}_{B}$--valued functions, which is precicely the description introduced in \cite{Ashtekar3}. 

A noteworthy feature of the limiting representations obtained in (\ref{polyA}) and (\ref{Wignerpi}), for quantum mechanics and the scalar field theory, respectively, is that the Wigner functions corresponding to the vacuum states occur to be invariant under translations. In the context of Wigner distributions, a vacuum state invariant under translations must satisfy, $U(u)*\rho_{0}=\rho_{0}$, where the star product was defined in (\ref{star}) \cite{Compean},\cite{Zachos}. A straightforward calculation shows that this takes place precisely when $1/d\to 0$ or the inverse of the covariance operator $C^{-1}\to 0$ weakly, depending the case considered. This implies that the non--regular representations associated to the polymer--like models is related with the gauge invariance of ground states, which corresponds to the claim made by a generalization of the Stone--von Neumann theorem depicted in \cite{Cavallaro}.

\section{Conclusions}
\label{sec:conclu}

In this paper, we showed that the non--regular polymer representation of the real--valued scalar field theory can be obtained by taking the limit of Gaussian measures in the functional Schr\"odinger representation. Particularly, we argued that the polymer Wigner functional corresponds to the product of polymer Wigner functions defined over the Bohr compactificaction of the real line. This implies that the limiting representation corresponds to the polymer representation derived by using algebraic methods such as the GNS construction, and the Fock quantization procedure. We must emphasize that the techniques developed here, in principle, can be applied to generic field theories. In particular we are interested on some issues on QFT such as the appearence of ultraviolet divergences. It should be possible to use the Wigner functionals, defined on quantum geometrical states, and the semiclassical methods established on deformation quantization to trace the manner in which the continuum limit generates divergences. Additionally, it is interesting to examine the quantization of the holonomy--flux algebra within this approach, since, precicely this instance leads to the mathematical underpinning of LQG. However, a more general analysis must be carried out in order to treat with diffeomorphism invariance, thereby, we intend to devote a forthcoming work to address these questions.

\section*{Acknowledgments}
The author would like to acknowledge financial support from CONACYT--Mexico
under project CB--2017--283838.

\section*{References}

\bibliographystyle{unsrt}

\end{document}